\documentclass[twocolumn]{pasj01}
\Received{$\langle$22-Feb-2019$\rangle$}
\Accepted{$\langle$18-Mar-2019$\rangle$}
\Published{$\langle$publication date$\rangle$}
\SetRunningHead{Astronomical Society of Japan}{Usage of \texttt{pasj00.cls}}
\usepackage{times}
\begin{document}

\title{Ground-based Pa$\alpha$ Narrow-band Imaging of Local Luminous Infrared Galaxies II: Bulge Structure And Star Formation Activity}

\author{
Ken~\textsc{Tateuchi},\altaffilmark{1,2}
Kentaro~\textsc{Motohara},\altaffilmark{2}
Masahiro~\textsc{Konishi},\altaffilmark{2}
Hidenori~\textsc{Takahashi},\altaffilmark{2}
Yutaro~\textsc{Kitagawa},\altaffilmark{2}
Natsuko~\textsc{Kato},\altaffilmark{2}
Soya~\textsc{Todo},\altaffilmark{2}
Shinya~\textsc{Komugi},\altaffilmark{3}
Ryou~\textsc{Ohsawa},\altaffilmark{2}
Mamoru~\textsc{Doi},\altaffilmark{2}
and
Yuzuru~\textsc{Yoshii}\altaffilmark{2,4}
}
\altaffiltext{1}{School of Law, Graduate Schools for Law and Politics, The University of Tokyo, 7-3-1 Hongo, Bunkyo-ku, Tokyo 113-0033, Japan}
\email{tateuchi@ioa.s.u-tokyo.ac.jp}
\altaffiltext{2}{Institute of Astronomy, The University of Tokyo, 2-21-1 Osawa, Mitaka, Tokyo 181-0015, Japan}
\altaffiltext{3}{Division of Liberal Arts, Kogakuin University, 2665-1, Hachioji, Tokyo 192-0015, Japan}
\altaffiltext{4}{Steward Observatory, University of Arizona, 933 North
Cherry Avenue, Rm. N204, Tucson, AZ 85721-0065, USA}


\maketitle

\begin{abstract}
We present properties of two types of bulges (classical and pseudo- bulges) in 
20 luminous infrared galaxies (LIRGs) observed in the near infrared of the $H$, 
$K_s$ and 1.91$\mu$m narrow-band targeting at the hydrogen Pa$\alpha$ emission 
line by the University of Tokyo Atacama Observatory (TAO) 1.0 m telescope. 
To classify the two types of bulges, we first perform a two-dimensional 
bulge-disk decomposition analysis in the $K_\mathrm{s}$-band images. The result 
shows a tentative bimodal distribution of S{\'e}rsic indices with a separation 
at $\log(n_b)\sim0.5$, which is consistent with that of classical and normal galaxies. 
We next measure extents of the distribution of star forming regions in Pa$\alpha$ 
emission line images, normalized with the size of the bulges, and find that 
they decrease with increasing S{\'e}rsic indices. These results suggest that 
star-forming galaxies with classical bulges have compact star forming regions 
concentrated within the bulges, while those with pseudobulges have extended star 
forming regions beyond the bulges, suggesting that there are different formation 
scenarios at work in classical and pseudobulges.
\end{abstract}

\section{Introduction}
Recent researches suggest that there are two types of bulges found at centers of galaxies, which are called ``classical bulges'' and ``pseudobulges'' (e.g., \cite{2004ARA&A..42..603K, 2005MNRAS.358.1477A, 2005RMxAC..23..101K, 2010ApJ...716..942F, 2009ASPC..419...31C, 2010ApJ...716..942F}). The classical bulges are characterized by a steep increase in density towards their centers. They are dynamically hot and supported by stellar velocity dispersion rather than their systemic rotational motion \citep{2008AJ....136..773F, 2013ApJ...772...36G}. In addition, they are red in color, old in population, and high in metallicity and in $\alpha$/Fe ratio \citep{2006MNRAS.371..583M, 2006A&A...457L...1Z, 2007A&A...467..123B, 2007A&A...465..799L, 2008AJ....136..367M}. Also, it is suggested that their positions in the fundamental plane of the sizes versus surface magnitudes form a sequence, which is smoothly connected to that of elliptical galaxies \citep{2002MNRAS.335..741F, 2006MNRAS.366..510T, 2007A&A...474..763J, 2007ApJ...657L..85D,2009MNRAS.393.1531G}. On the other hand, the pseudobulges are dynamically cold, supported by systemic rotation, and have flat shapes resembling those of exponential disks. Also, they can be identified as outliers in the fundamental plane \citep{2009MNRAS.393.1531G} and generally have younger stellar populations than those of classical bulges.
\citet{2004ARA&A..42..603K} have classified 75 normal galaxies in the $V$-band by eye, and find that 28.5\% of their sample have classical bulges and 71.5\% have pseudobulges. Using this sample, \citet{2008AJ....136..773F} show that a distribution of $n_b$, that is Sersic indicies　of bulges, is bimodal, where the pseudobulges have $n_b$ $<$ 2.2 and the classical bulges $n_b$ $\geq$ 2.2, and that types of bulges can be distinguished by the S{\'e}rsic index.

Theoretically, the classical bulges are thought to be formed by major merger processes (e.g., \cite{2004ARA&A..42..603K, 2006MNRAS.369..625N, 2010ApJ...715..202H}), or by collapse of giant clumps in primordial disks \citep{2008ApJ...688...67E} to remove their angular momentum. Elliptical galaxies are also expected to be formed by similar processes but through dissipationless (dry) merger, where their progenitors have little gas content. (e.g., \cite{2004ARA&A..42..603K,2009ApJS..182..216K}). On the other hand, the pseudobulges are suggested to be built by a secular evolution (e.g., \cite{2004ARA&A..42..603K}). \citet{2013MNRAS.428..718O} shows by numerical cosmological simulations that some pseudobulges are formed through starburst events even at redshifts of two, which may be an extreme example of the secular evolution.

Furthermore, it is speculated that the formation of the two types of bulges has a deep relationship with growth of super-massive black holes (SMBHs). \citet{1998AJ....115.2285M} find a correlation between mass of a central SMBH ($M_{\bullet}$) and velocity dispersion of a host ($\sigma$) in 32 dust-free nearby galaxies, which is followed-up by various observations. This is now known as ``$M- \sigma$'' relation, suggesting that the SMBHs and their hosts co-evolve (e.g., \cite{2001AIPC..586..363K,2000ApJ...539L...9F,2000ApJ...539L..13G,2002ApJ...574..740T,2009ApJ...698..198G}). \citet{2011Natur.469..374K} also find that the $M-\sigma$ relation of classical bulges has a tight correlation, while that of pseudobulges does not. They suggest from this discovery that there are two fundamentally different feeding mechanisms for SMBHs; in a classical bulge, gas is fed to a SMBH through a rapid gas inflow process towards the center of the galaxy, which also result in the stellar mass accumulation of the bulge, while in a pseudobulge gas feeding to a SMBH in a pseudobulge which is controlled by a local process working stochastically at a central $\sim$ 10$^2$ pc region, and is independent or has a weak connection with the growth of the bulge.
Thus, the two types of bulges may have different formation processes, where classical bulges are formed by drastic external factors such as major merger processes similar to those of elliptical galaxies, while pseudobulges evolve more secularly. However, there are few observational studies to verify the formation scenario.

To understand the formation process of bulges, we focus on LIRGs (Luminous Infrared Galaxies; $L_{\mathrm{IR}}$ $\equiv$ 10$^{11}$--10$^{12}$ $L_{\solar}$) 
in the local universe. They are ideal laboratories for studying bulge formation, because half of them are non-irregular galaxies \citep{2006ApJ...649..722W} whose bulges can be evaluated easily, and might be current formation sites of bulges with ongoing starbursts. Although the LIRGs may provide a direct way to understand a detailed mechanism of the bulge formation, they are affected by a large amount of dust produced by their active star formations, and optical hydrogen recombination lines such as H$\alpha$ and H$\beta$, which are direct probes of star formation activities, are easily attenuated by the dust. We therefore focused on the hydrogen Pa$\alpha$ line (1.8751 $\mu$m) which is relatively insensitive to dust extinction ($A_{\mathrm{Pa}\alpha}$/$A_{\mathrm{H}\alpha} = 5.68$) and the strongest emission line in the near-infrared ($\lambda$ = 0.9--2.5 $\mu$m) wavelengths.There are several Pa$\alpha$ narrow-band imaging studies of nearby LIRGs by the Near Infrared Camera and Multi-object Spectrometer (NICMOS) on $Hubble\ Space\ Telescope$ ($HST$) (e.g., \cite{2006ApJ...650..835A,2007ApJ...666..870C,2007ApJ...671..333K,2009ApJ...692..556R}). However, some researchers have pointed out that HST/NICMOS may be insensitive to diffuse Pa$\alpha$ emission due to its intrinsic high angular resolution (e.g., \cite{2006ApJ...650..835A,2007ApJ...666..870C,2007ApJ...671..333K,2009ApJ...692..556R}). On the other hand, due to poor atmospheric transmission around the Pa$\alpha$ wavelength, it is difficult to carry out Pa$\alpha$ observation from the ground.


We have carried out Pa$\alpha$ narrow-band imaging survey of local LIRGs \citep{2012PKAS...27..297T,2013ASPC..476..301T,2015ApJS..217....1T} with Atacama Near InfraRed camera (ANIR; \cite{2008SPIE.7014E..94M,2010SPIE.7735E.120M,2015PASJ..tmp..154K}) on the University of Tokyo Atacama Observatory 1.0 m telescope (miniTAO; \cite{2010SPIE.7733E.163M,2010SPIE.7733E..08Y}) installed at the summit of Co. Chajnantor (5640m altitude) in northern Chile. Thanks to the high altitude and the extremely low water vapor of the site, we can stably observe Pa$\alpha$ emission line even from the ground \citep{2010SPIE.7735E.120M}. 
In this paper, we evaluate the properties of bulges and star forming regions in the LIRGs using the survey data to probe formation process of classical and pseudobulges.

This paper is organized as follows. We introduce the survey data and describe a selection method of non-irregular galaxies and a method of their bulge-disk decomposition in $\S$ 2. In $\S$ 3, we present the results and discuss the relationship between the types of bulges and the size of star forming regions, and summarize them in $\S$ 4.

Throughout this paper, we use a $\Lambda$-CDM cosmology with $\Omega_{m}$ = 0.3, $\Omega_{\Lambda}$ = 0.7, and $H_0$ = 70 $\rm km$ $\rm s^{-1}$ $\rm Mpc^{-1}$.

\section{Sample and Data Analysis}

\begin{figure}
 \begin{center}
  \includegraphics[width=8cm]{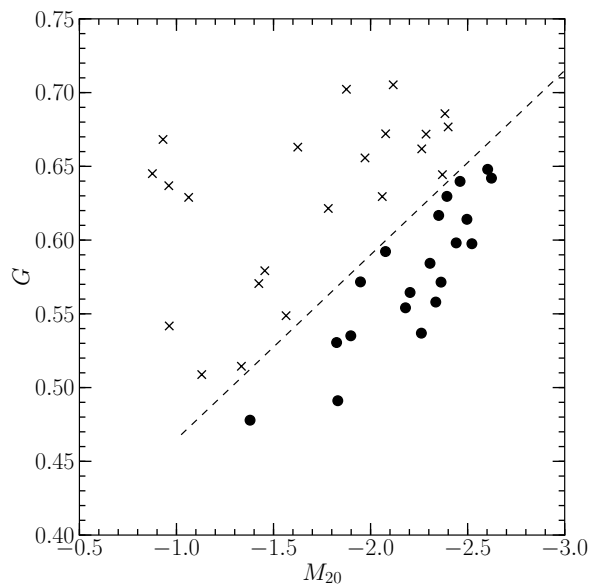} 
\end{center}
\caption{Diagram of Gini coefficients ($G$) and second-order moments of the brightest 20\% of flux of galaxies ($M_{20}$) measured in the $K_\mathrm{s}$-band images. The dashed line is the threshold between irregulars (crosses) and non-irregulars (circles) defined by \citet{2004AJ....128..163L}. 
\label{fig1}}
\end{figure}

\subsection{Sample}
We use $K_\mathrm{s}$-band and Pa$\alpha$ line images of LIRGs obtained by \citet{2015ApJS..217....1T}, consisting of 38 galaxies selected from the Infrared Astronomical Satellite ($IRAS$) Revised Bright Galaxy Sample (RBGS : \citep{2003AJ....126.1607S}) with the bolometric infrared luminosity ($L_{\mathrm{IR}}$) ranging between $4.5\times10^{10}$ and $6.5\times10^{11}$ $L_{\odot}$. Their recession velocities are 2800 -- 8100 km s$^{-1}$ , corresponding to distances of 46.6 -- 109.6 Mpc.

\def\arraystretch{1.25}
\begin{longtable}{*{10}{c}}
\multicolumn{10}{c}{}\\
\caption{Sample of Non-irregular Luminous Infrared Galaxies\label{table_bulgesample}}
\hline
\hline
Galaxy & Hubble & Spectral & & & $R_\mathrm{sf}$ & $R_b$ & & & \\
Name & Type & Class & $G$ & $M_{20}$ & (kpc) & (kpc) & $n_b$ & $n_b^{HST} $ & $\log(B/T)$ \\
(1) & (2) & (3) & (4) & (5) & (6) & (7) & (8) & (9) & (10)\\
\hline
\endhead
\endfoot
\multicolumn{10}{@{}l}{\rlap{\parbox[t]{.80\textwidth}{\footnotesize
Column (1): Galaxy names. Column (2): Hubble type obtained by HyperLeda database (http://leda.univ-lyon1.fr). Column (3): Classification by optical spectroscopic observation taken from NASA/IPAC Extragalactic Database (NED) (https://ned.ipac.caltech.edu/). Sy1: Seyfert1, Sy2: Seyfert2, LINER: LINER, H{\sc ii}: H{\sc ii} region, and N: unknown. Column (4): Gini coefficient. Column (5): Second-order moment of the brightest 20\% of flux of a galaxy. Column (6): Effective radius of distribution of the Pa$\alpha$ emission. Column (7): Effective radius of a bulge measured in the $K_\mathrm{s}$-band. Column (8): S{\'e}rsic index of a bulge measured in the $K_\mathrm{s}$-band image. Column (9): S{\'e}rsic index of a bulge measured in $HST$ NICMOS $F160W$ image. Column (10): A ratio of bulge to total (bulge $+$ disk) luminosity. }}}
\endlastfoot
NGC 23\dotfill						&	1	&	H{\sc ii}	&	0.63	&	--2.39	&	1.59	&	0.82 	&	1.9 $\pm$ 0.03	&	1.9	&	--0.4	\\
NGC 232\dotfill						& 	1	&	Sy2	&	0.59	&	--2.08	&	0.61	&	1.10	&	2.3 $\pm$ 0.03	&	--	&	--0.2	\\
IRAS F02437$+$2122\dotfill			&	--5	&	LINER	&	0.61	&	--2.50	&	0.23	&	0.55	&	4.6 $\pm$ 0.26	&	--	&	--0.4	\\
NGC 1720\dotfill					&	2	&	N	&	0.65	&	--2.60	&	0.46	&	0.45	&	1.8 $\pm$ 0.02	&	--	&	--0.7	\\
ESO 557-G002\dotfill				&	4	&	H{\sc ii}	&	0.54	&	--2.26	&	1.34	&	73.82	&	19.9 $\pm$ 2.1	&	--	&	--0.2	\\
ESO 320-G030\dotfill				&	1	&	H{\sc ii}	&	0.54	&	--1.90	&	0.84	&	0.47	&	1.2 $\pm$ 0.02	&	--	&	--0.4	\\
MCG $-$03-34-064\dotfill			&	--2	&	Sy1	&	0.62	&	--2.35	&	0.25	&	0.36	&	1.6 $\pm$ 0.03	&	--	&	--0.4	\\
NGC 5135\dotfill					&	2	&	Sy2	&	0.60	&	--2.52	&	1.63	&	0.70	&	1.0 $\pm$ 0.02	&	1.0	&	--0.5	\\
NGC 5257\dotfill					&	3	&	H{\sc ii}	&	0.48	&	--1.38	&	5.33	&	0.22	&	1.6 $\pm$ 0.2	&	2.0	&	--1.4	\\
IC 4687\dotfill						&	3	&	H{\sc ii}	&	0.57	&	--1.95	&	1.27	&	5.05	&	6.2 $\pm$ 0.31	&	--	&	--0.6	\\
ESO 339-G011\dotfill				&	3	&	Sy2	&	0.56	&	--2.34	&	3.12	&	0.38	&	0.6 $\pm$ 0.02	&	--	&	--0.8	\\
IC 5063\dotfill						&	0	&	Sy2	&	0.58	&	--2.31	&	0.44	&	4.05	&	5.7 $\pm$ 0.08	&	--	&	--0.2	\\
NGC 7130\dotfill					&	1	&	LINER	&	0.53	&	--1.82	&	2.22	&	0.62	&	6.1 $\pm$ 0.32	&	--	&	--0.5	\\
IC 5179\dotfill						&	4	&	H{\sc ii}	&	0.49	&	--1.83	&	4.94	&	1.62	&	6.5 $\pm$ 0.43	&	--	&	--0.8	\\
ESO 534-G009\dotfill				&	2	&	LINER	&	0.56	&	--2.20	&	0.18	&	1.30	&	2.6 $\pm$ 0.02	&	--	&	--0.2	\\
CGCG 453-062\dotfill				&	2	&	LINER	&	0.55	&	--2.18	&	2.49	&	0.23	&	1.2 $\pm$ 0.15	&	--	&	--1.0	\\
NGC 7591\dotfill					&	4	&	LINER	&	0.64	&	--2.46	&	0.62	&	3.06	&	4.0 $\pm$ 0.04	&	--	&	--0.2	\\
MCG $-$01-60-022\dotfill			&	7	&	H{\sc ii}	&	0.60	&	--2.44	&	3.19	&	0.26	&	1.5 $\pm$ 0.08	&	--	&	--0.7	\\
NGC 7771\dotfill					&	1	&	H{\sc ii}	&	0.57	&	--2.36	&	11.76	&	0.86	&	0.4 $\pm$ 0.01	&	--	&	--0.9	\\
UGC 12915\dotfill					&	5	&	LINER	&	0.64	&	--2.62	&	7.86	&	0.54	&	1.0 $\pm$ 0.01	&	--	&	--0.6 \\
\hline
\end{longtable}
\def\arraystretch{1}

\subsection{Identification of Non-irregular Galaxies}
We determined whether a galaxy is irregular or not by a combination of a Gini coefficient ($G$) and a second-order moment ($M_{20}$) of its surface-brightness distribution, which are nonparametric methods for quantifying galaxy morphology \citep{2004AJ....128..163L}. $G$ is a statistic based on the Lorenz curve of fluxes per pixel in a galaxy and represents the relative distribution of pixels covering the galaxy, while $M_{20}$ is a normalized second-order moment of pixel flux values which is measured to be the brightest 20\% of the fluxes of a galaxy. Figure \ref{fig1} shows the $M_{20}$--$G$ plot measured in the $K_\mathrm{s}$-band. These parameters are affected by spatial resolution and noise level of the image. \citet{2004AJ....128..163L} find that they are reliable if the averaged signal-to-noise ratio ($\langle S/N \rangle$) per pixel of an image is larger than 2. Also, they find that $M_{20}$ tends to have a systematic offset greater than $\sim$ 15\% when the spatial resolution is larger than 500 pc, as a core of a galaxy becomes unresolved, while $G$ is relatively stable by decreasing the resolution until the smallest resolved detail is 1000 pc. In our sample, the resolution is 180--400 pc corresponding to typical seeing size of 0$\farcs$8 and $\langle S/N \rangle$ is over five, which are enough to evaluate the morphology of the galaxies by $G$ and $M_{20}$. We adopt a criterion defined by \citet{2004AJ....128..163L} to distinguish irregular galaxies from the others on the $G$--$M_{20}$ diagram, as shown in Figure \ref{fig1}. We thus selected 20 non-irregular galaxies out of 38 galaxies for further analysis. The fraction of non-irregular galaxies is consistent with that of \citet{2006ApJ...649..722W} claiming that half of LIRGs in the local universe are classified as non-irregular galaxies.

\subsection{Bulge-Disk Decomposition}\label{sec:Bulge-Disk Decompositions}
In UV or optical observations, it is difficult to evaluate a bulge structure of a LIRG because it is obscured by dust \citep{2010ApJ...716..942F}. Instead, as $K_\mathrm{s}$-band observations are less affected by dust extinction, they are expected to reflect the properties of the stellar mass distribution \citep{2003ApJS..149..289B}. To evaluate the properties of bulges of LIRGs, we therefore evaluate S{\'e}rsic indices of the bulges by two-dimensional bulge-disk decomposition analysis in the $K_\mathrm{s}$-band images using the software GALFIT \citep{2002AJ....124..266P, 2010AJ....139.2097P}. We fit the surface brightness of each galaxy with a combination of a S{\'e}rsic profile as a bulge component and an exponential profile as a disk component,
\begin{eqnarray}
I(r) = I_b \exp{[-(r/R_b)^{1/n_b}]} + I_d \exp{[-(r/R_d)]},
\end{eqnarray}
where $r$ is a distance from the galactic center, $I_b$, $R_b$, $n_b$ are a central surface brightness, an effective radius, and a Sersic index of the bulge, respectively, and $I_d$ and $R_d$ are a central surface brightness and a scale length of the disk, respectively. We assume that the peak of flux in the $K_\mathrm{s}$-band image is the position of the galactic center, and use it as an initial input parameter. A PSF is measured by stacking stars detected in the $K_\mathrm{s}$-band image. 

The results of the bulge-disk decomposition are shown in Figure \ref{fig:profiles}. For each galaxy, the left panel shows a $K_\mathrm{s}$-band image observed by miniTAO/ANIR, the second from the left a model galaxy composed of the best-fit S{\'e}rsic and exponential components, and the third from the left a residual image created by subtracting the model galaxy from the $K_\mathrm{s}$-band image. The residual structure of spiral arms and bars can be seen in the residual image for most of the sample. The right panel shows a radial profile of the surface brightness with model profiles, cut along the major axis of the galaxy.

There is a possibility that the limited spatial resolution (0$\farcs$8 seeing) in our sample may miss complex structures in a bulge such as double nuclei, and biases the results of the bulge-disk decomposition \citep[e.g.]{2011AJ....141..100H,2016A&A...591A...1P}. Therefore, we evaluate S{\'e}rsic indices of the bulges having $HST$ NICMOS $F160W$ data with the same method. These results are almost the same as our results listed in Figure \ref{fig:profiles}. 

\section{Result and Discussion}
\subsection{Properties of Bulges}
\subsubsection{S{\'e}rsic Indices of Bulges}
\begin{figure}
\begin{center}
\includegraphics[width=8cm]{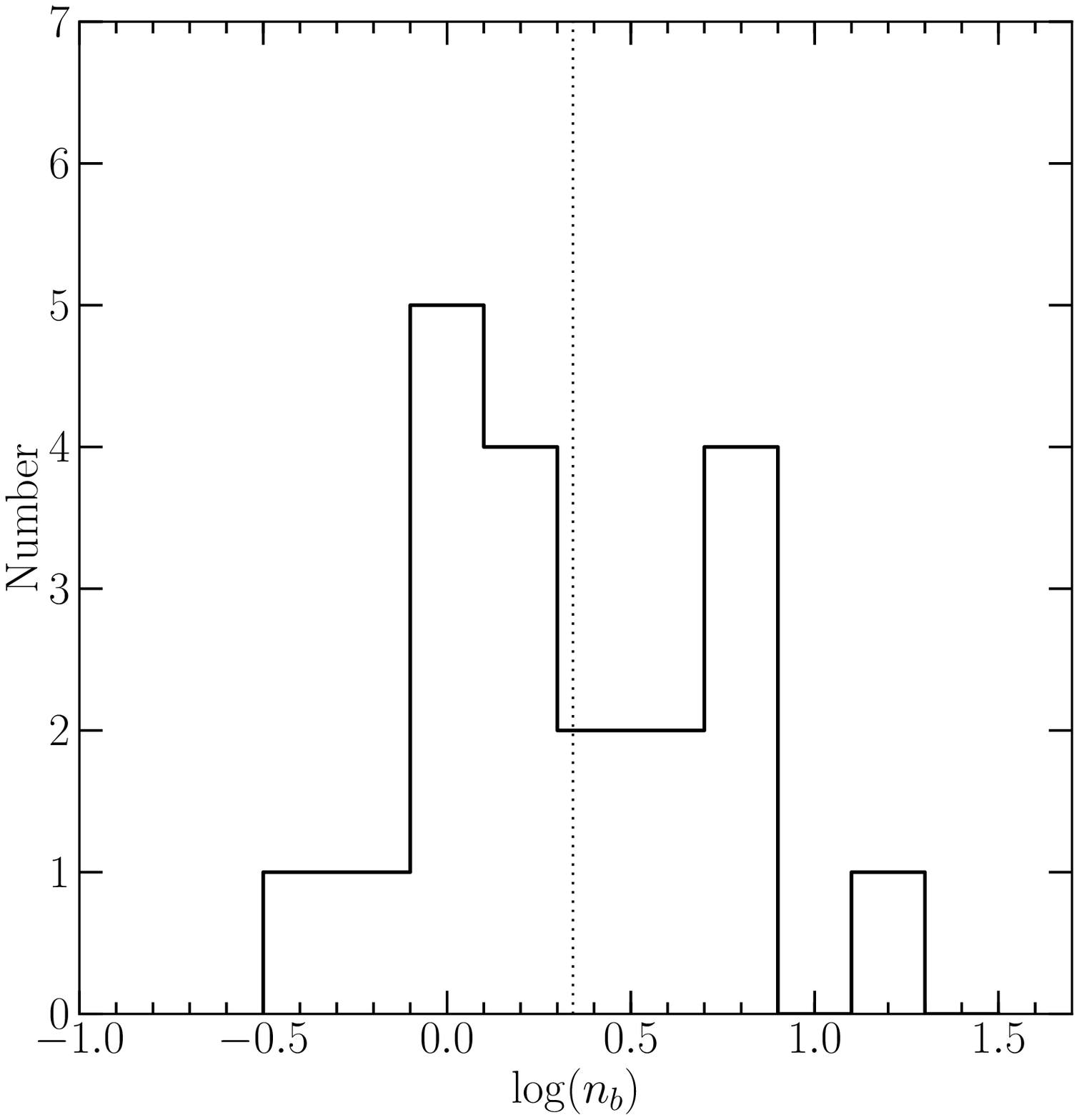} 
\end{center}
\caption{Distribution of the S{\'e}rsic index of the bulges measured in the $K_\mathrm{s}$-band. The dotted line represents the border of the classical and pseudobulge classification (classocal bulge ($n_b\geq2.2$), pseudobulges ($n_b<2.2$))} shown by $V$-band study \citep{2008AJ....136..773F}. \label{fig:figure_nb_histgram1}
\end{figure}

\citet{2008AJ....136..773F} measure S{\'e}rsic indices of normal galaxies in the $V$-band and show that classical and pseudobulges show a bimodal distribution of Sersic indices separated at $n_b$ $=$ 2.2 ( $\log(n_b)=0.3$ ). They also claim that the S{\'e}rsic indices measured in the $V$-band and in the $H$-band have almost the same value, suggesting that the criteria can also be used in the $K_\mathrm{s}$-band. In Figure \ref{fig:figure_nb_histgram1}, we show a histogram of the S{\'e}rsic indices of the bulges measured in the $K_\mathrm{s}$-band. Although it is inconclusive statistically, it shows a bimodal distribution with a separation at $\log(n_b)\sim0.5$, which is consistent with the result of \citet{2008AJ....136..773F}, although the bimodality is statistically tentative. If we follow the criteria of \citet{2008AJ....136..773F} and separate classical and pseudobulges at $n_b$ $=$ 2.2 ( $\log(n_b)=0.3$ ), 45\% (9/20) of our sample have classical bulges and 55\% (11/20) pseudobulges. Although the volume of our sample is not large enough statistically to make a firm conclusion and the difference in fraction of bulges and pseudobulges is not established at a significant level, this suggests that the fraction of classical bulges in LIRGs is higher than that in normal galaxies of 28.5\% \citep{2004ARA&A..42..603K}. As classical bulges are expected to be formed through major merger processes (e.g., \citep{2004ARA&A..42..603K,2008AJ....136..773F}), this is consistent with previous works that LIRGs may have experienced more major merger events than normal galaxies (e.g., \cite{1996ARA&A..34..749S,2006ApJ...649..722W}).

\begin{figure}
\begin{center}
\includegraphics[width=8cm]{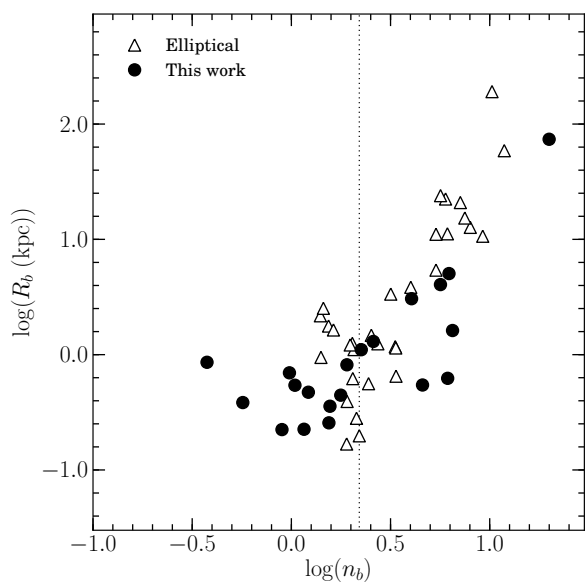} 
\end{center}
\caption{S{\'e}rsic indices and effective radii of the bulges of our sample of galaxies measured in the $K_\mathrm{s}$-bands. Triangles show elliptical galaxies taken from \citet{2009ApJS..182..216K}. The dotted line represents the border of classical and pseudo-bulges shown in Figure 2. \label{fig:figure_nb_histgram2}}
\end{figure}

\subsubsection{Size and Luminositiy of Bulges}
Figure \ref{fig:figure_nb_histgram2} shows a relationship between the S{\'e}rsic indices and the effective radii of the bulges. The effective radii of the pseudobulges remain constant of the order of sub-kiloparsec, while those of the classical bulges increase with the S{\'e}rsic indices, which is consistent with the distribution of normal galaxies \citep{2008AJ....136..773F}. 

\begin{figure}[t]
\begin{center}
\includegraphics[width=8cm]{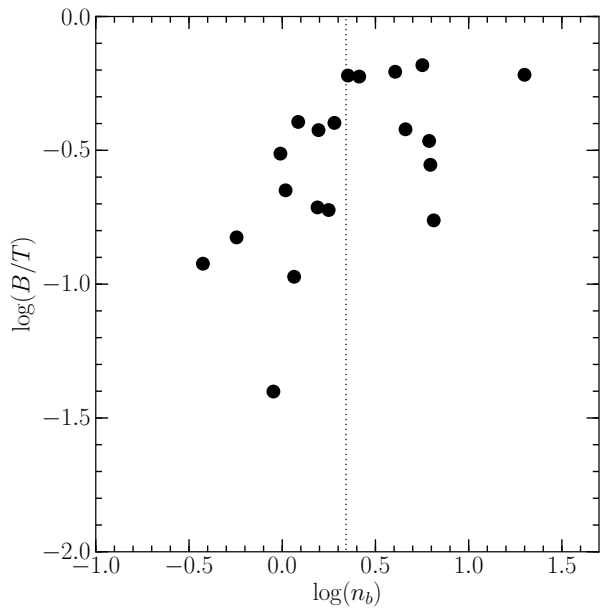} 
\end{center}
\caption{Relationship between S{\'e}rsic indices and bulge-to-total luminosity ratio ($B/T$) obtained from a model profile by the bulge-disk decomposition in the $K_\mathrm{s}$-band image. The dotted line shows the border of classical and pseudo- bulges shown in Figure 2. \label{fig:plot_sersic_ktot}}
\end{figure}

Figure \ref{fig:plot_sersic_ktot} shows a relationship between the S{\'e}rsic indices and $B/T$ ratios, which are obtained from a surface brightness profile fitting in Section 2.3. It suggests that the $B/T$ ratios increase with the S{\'e}rsic indices, again consistent with the properties of normal galaxies \citep{2008AJ....136..773F}.

The results presented in this section suggest that the properties of bulges in LIRGs are similar to those in normal galaxies except for the fraction of classical and pseudobulges.

\subsection{Distribution of Star Forming Regions}
To understand a relationship between star formation activities and types of bulges, we derive the extent of the distribution of star forming regions in the Pa$\alpha$ narrow-band images. The extent is defined to be a half light radius with ellipticity and position angle the same as those of the bulges is in the $K_\mathrm{s}$-band image in a Pa$\alpha$ emission line image ($R_\mathrm{sf}$), and shown in the fourth panels from the left in Figure 6 and Table \ref{table_bulgesample}. From this sample, we remove $IRAS$ F02437$+$2122 whose $R_\mathrm{sf}$ is almost equal to the size of the PSF and $\mathrm{MCG}\ -03$-34-064 which is a Seyfert $1$, whose concentrated Pa$\alpha$ emission may be strongly contaminated by the AGN. Accordingly, there are 18 galaxies of which we can evaluate $R_\mathrm{sf}$. The effect of AGNs for these galaxies is not so large considering the low effect of AGNs for bolometric luminosities of LIRGs ($\sim$5\%; \cite{2012ApJ...744....2A}) and Pa$\alpha$ flux correlates with far infrared flux well \citep{2015ApJS..217....1T}.

\begin{figure}[h]
\begin{center}
\includegraphics[width=8cm]{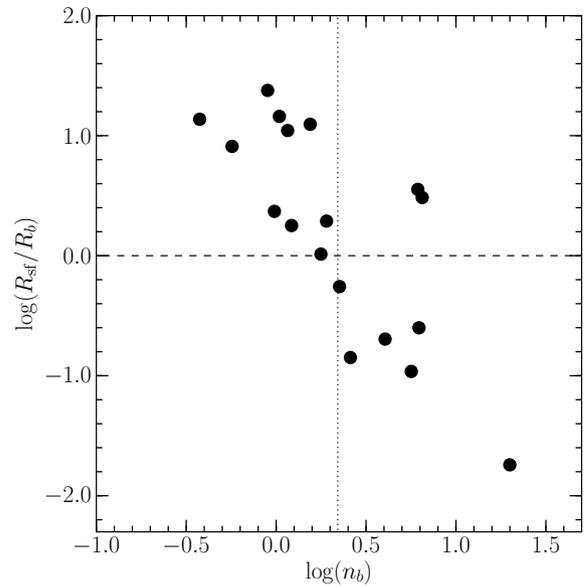} 
\end{center}
\caption{Comparison between the S{\'e}rsic index of bulge ($n_b$) and the normalized size of the star forming regions ($R_\mathrm{sf}/R_b$). The horizontal dashed line represents that $R_\mathrm{sf}$ is equal to  $R_b$, which means the size of the star forming regions is the same as that of the bulge. The dotted line shows the border of classical and pseudo- bulges shown in Figure 2. The high S{\'e}rsic indices of bulges represent compact star forming regions while low S{\'e}rsic indices of bulges represent extended star forming regions. \label{fig:plot_sersicr50.eps}}
\end{figure}

We normalize $R_\mathrm{sf}$ by the effective radii of the bulges ($R_b$) and show a relationship between the S{\'e}rsic indices $n_b$ and $R_\mathrm{sf}/R_b$ in Figure \ref{fig:plot_sersicr50.eps}. It can be seen that $R_\mathrm{sf}/R_b$ decreases with increasing $n_b$ and that $R_\mathrm{sf}$ becomes equal to $R_b$ at $n_b\sim 2.2$, which is the point to separate classical and pseudobulges. These results suggest that starburst galaxies with classical bulges have compact star forming regions concentrated within the bulges, while those with pseudobulges have extended star forming regions beyond the bulge, possibly along their spiral arms.

Classical bulges are expected to have experienced major merger events to lose angular momentum and have similar properties to elliptical galaxies \citep{2004ARA&A..42..603K}. Theoretically, gas tends to be concentrated at a central region of a galaxy after a major merger event \citep{2006MNRAS.369..625N,2013MNRAS.430.1901H}. Also, there is a simulation which forms a classical bulge with inflow of cold gas toward a center of a galaxy \citep{2012MNRAS.423.1544S}, inducing extreme star formation at the central region of the galaxy. Our result, in which the star formation is concentrated at the centers in the classical bulges, is consistent with such theoretical simulations. On the other hand, pseudobulges are expected to be formed by minor merger events or secular evolution in which galaxies evolve over a long time without any strong interaction (e.g., \cite{2004ARA&A..42..603K}). In these processes, star forming activities are expected to occur along their arms and bar-ends and spread beyond the size of the bulge. Our result, where the galaxies with pseudobulges have extended star forming regions beyond their bulges, is consistent with such pictures.

There is a question whether the observed concentrated star formation in classical bulges is not directly connected with the major merger event that formed the bulges, especially because we have selected non-irregular galaxies, which show little traces of mergers, as the sample. Numerical simulations show that enhanced star formation is expected to peak after the strong morphological disturbances of the first encounter and lasts significantly longer than the merger phase (about 500 Myr after coalescence phase), during which the star formation is enhanced by more than 10 times (e.g., \cite{2008MNRAS.391.1137L,2010ApJ...720L.149T}). Considering that typical relaxation time scale of stars in a merger remnant is a few million years after the coalescence phase \citep{2013MNRAS.430.1901H}, our suggestion where classical bulges of LIRGs are formed in recent major merger events which also induce the current star formations, does not contradict with their non-irregular morphologies.

\subsection{Growth of SMBHs and Bulges}
Our results that classical bulges have compact star forming regions and pseudobulges have extended star forming regions suggests that growth of the classical bulges results from a large amount of gas falling into the centers of the galaxies via rapid feeding with major merger processes, which directly impact the growth of SMBHs. On the other hand, pseudobulges are suggested to grow together with disk components via secular evolution, where it is difficult for molecular gas to fall inwards toward the centers of the galaxies, and star formation activities are not always connected with the growth of SMBHs.
This picture supports the hypothesis of \citet{2011Natur.469..374K} that there are two different modes of a black hole feeding mechanism, where the growth of black holes and classical bulges are controlled by the same global process of rapid gas falling with major merging, while that of black holes and pseudobulges are independent or have a weak connection with the secular evolution.

\section{Summary}
In this paper, we have investigated properties of classical and pseudobulges in local LIRGs, which are considered to be a forming site of bulge structures, using the $K_\mathrm{s}$-band and Pa$\alpha$ emission line images observed using ANIR on the minTAO 1.0 m telescope. To classify the type of bulges, we perform a two-demensional bulge-disk decomposition analysis in the $K_\mathrm{s}$-band images. The result shows a tentative bimodal distribution of  S{\'e}rsic indices ($n_b$) with a separation at $\log(n_b)\sim0.5$, which is consistent with that of local normal galaxies. Also, the extent of the distribution normalized by the sizes of bulges ($R_\mathrm{sf}/R_\mathrm{b}$) in Pa$\alpha$ emission line images shows a negative correlation against $n_b$. This result means that galaxies with classical bulges have compact star forming regions concentrated within the bulges, while those with pseudobulges have extended star forming regions beyond the bulges, suggesting that there are different formation scenarios at work in classical and pseudobulges.




\begin{figure*}
\begin{center}
\includegraphics[width=16cm]{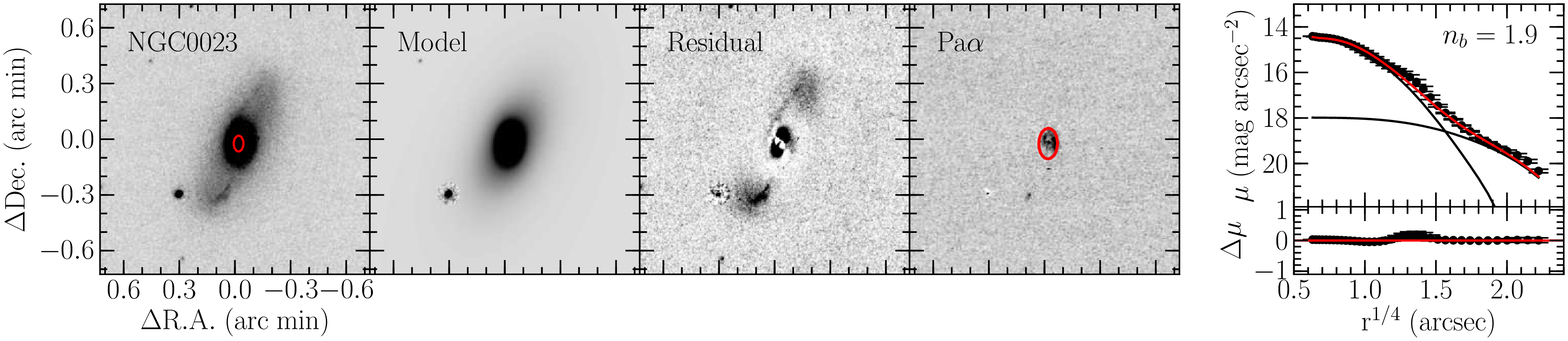}\\
\includegraphics[width=16cm]{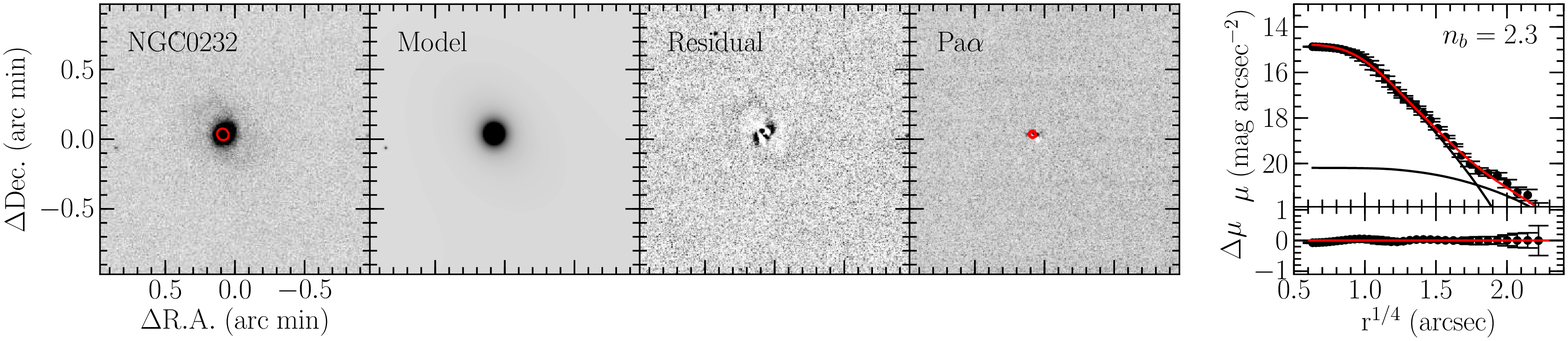}\\
\includegraphics[width=16cm]{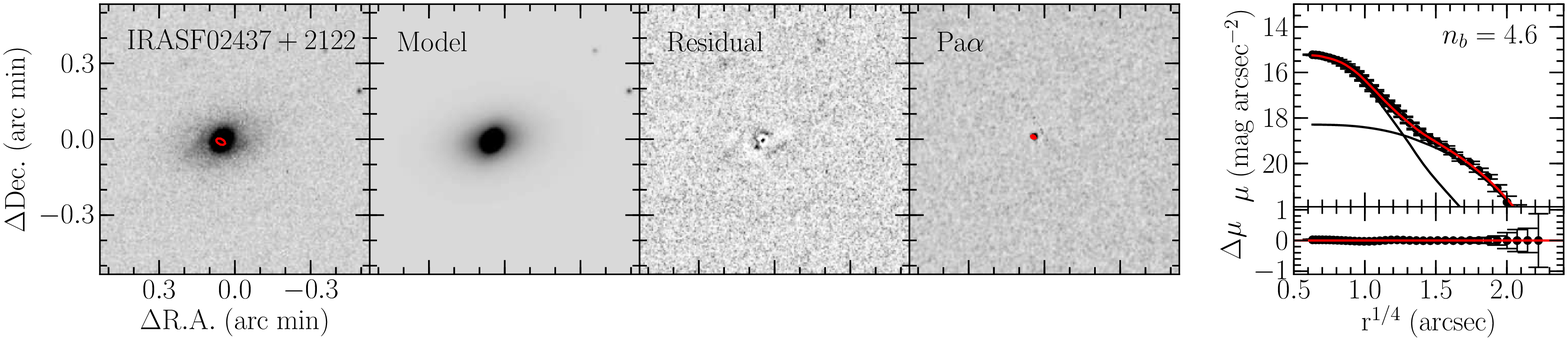}\\
\includegraphics[width=16cm]{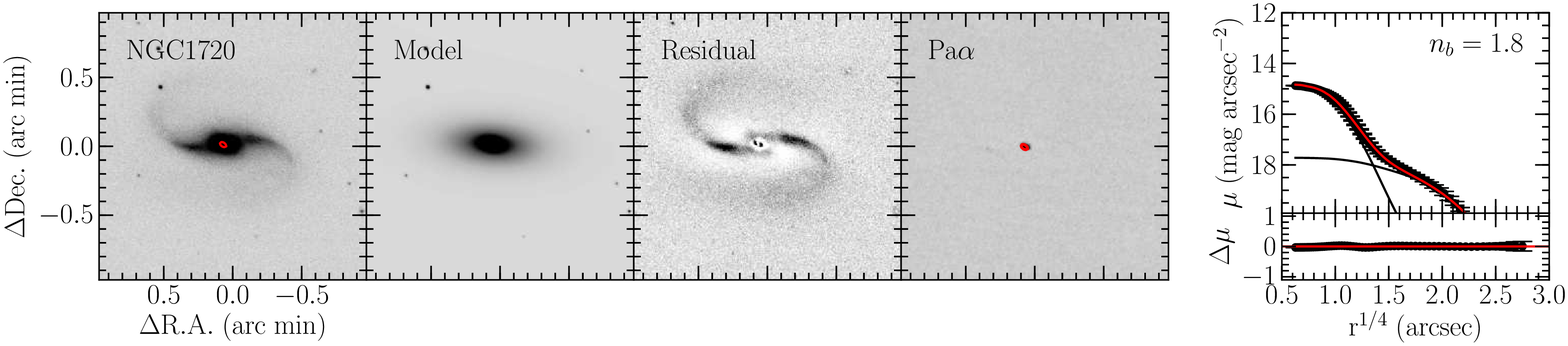}\\
\includegraphics[width=16cm]{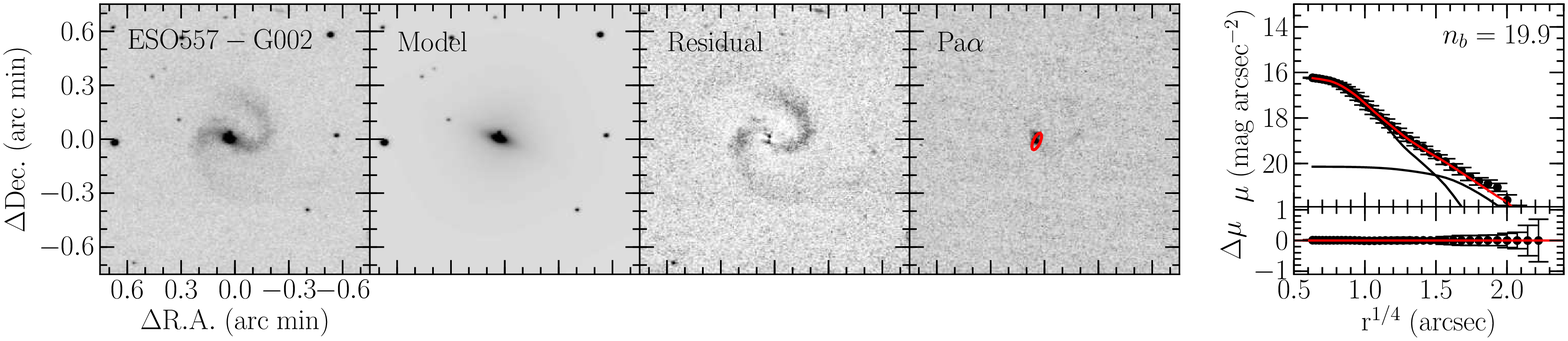}\\
\includegraphics[width=16cm]{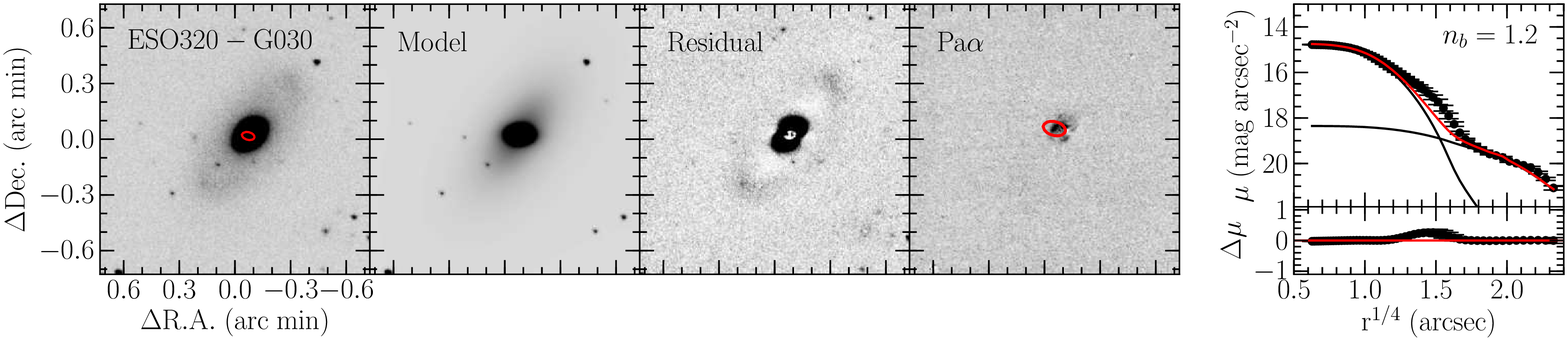}\\
\end{center}
\caption{Results of bulge-disk decomposition. For each galaxy, the $K_\mathrm{s}$-band image, a model image with best-fit bulge and disk components, a residual image after subtracting best-fit model image from the $K_\mathrm{s}$-band image and one-dimensional surface brightness profiles with residuals after fitting of the galaxy are shown, from left to right. Filled circles in the right panel are raw profile measured in the $K_\mathrm{s}$-band image with photometric error bars, solid lines are bulge and disk model profiles, and red solid line is the bulge-disk combined model profile. The red circles in the left and right panel are effective radius of the bulge and Pa$\alpha$ emission line regions, respectively. \label{fig:profiles}}
\end{figure*}

\addtocounter{figure}{-1}
\begin{figure*}
\begin{center}
\includegraphics[width=16cm]{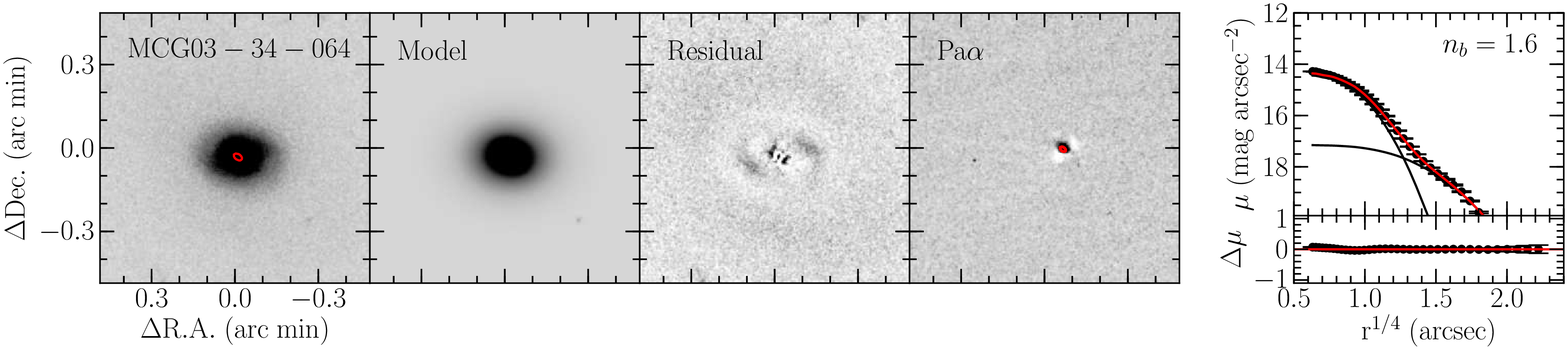}\\
\includegraphics[width=16cm]{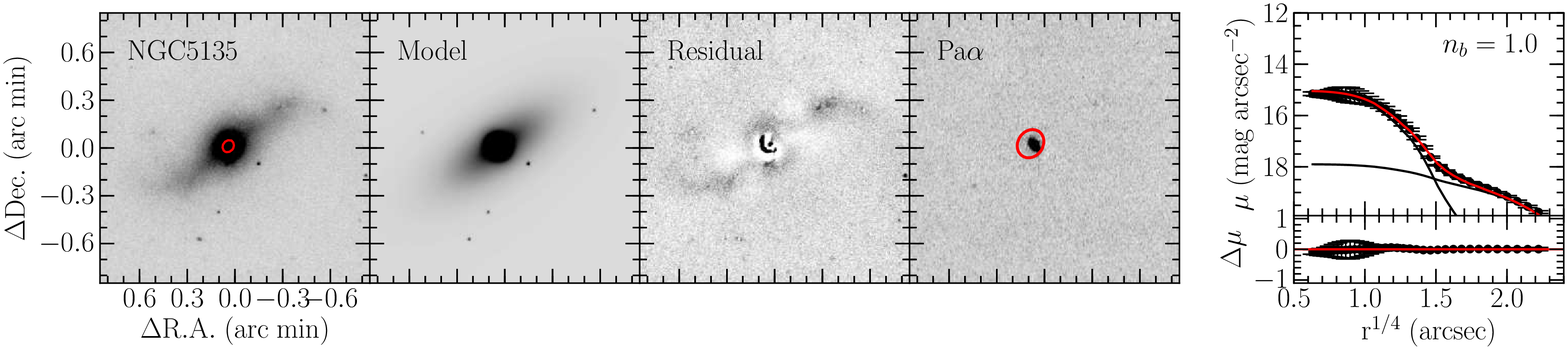}\\
\includegraphics[width=16cm]{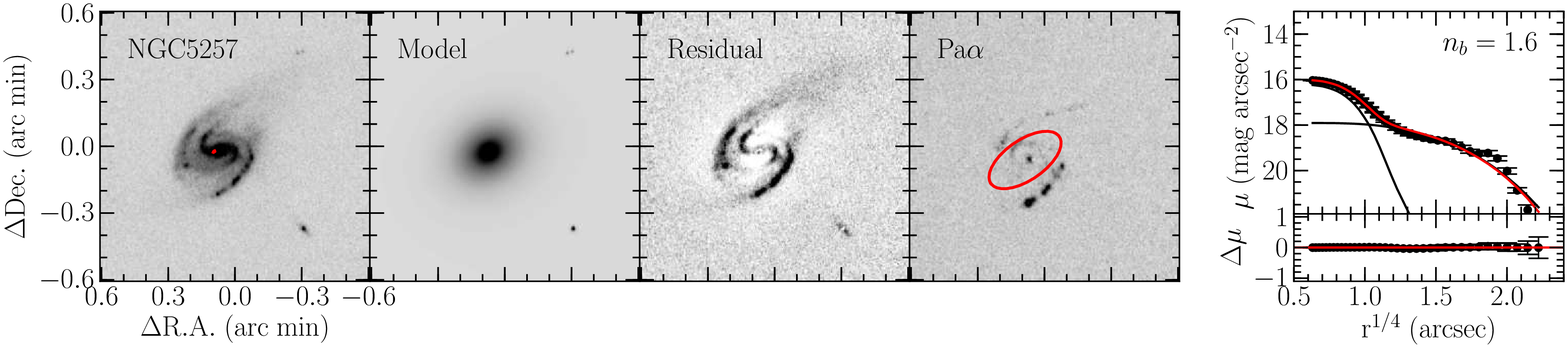}\\
\includegraphics[width=16cm]{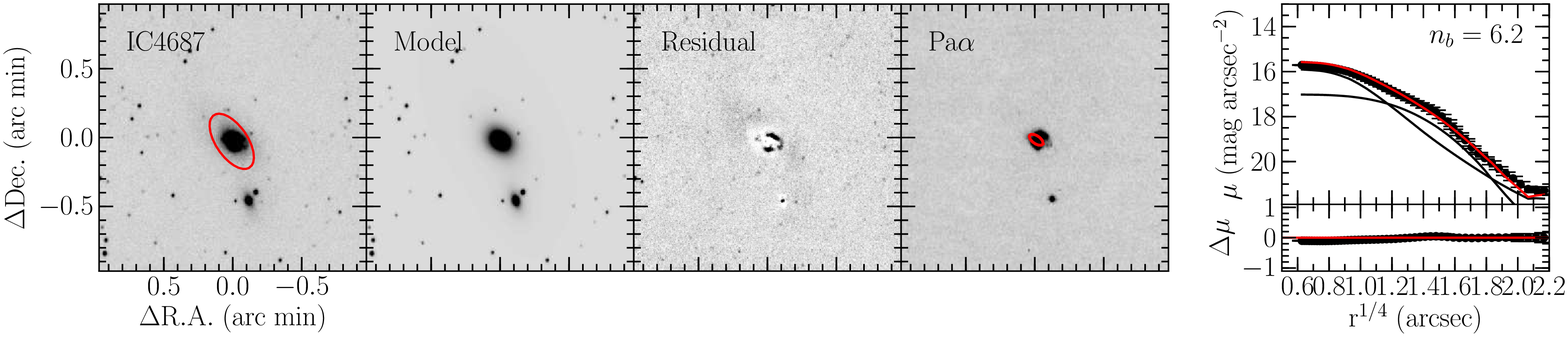}\\
\includegraphics[width=16cm]{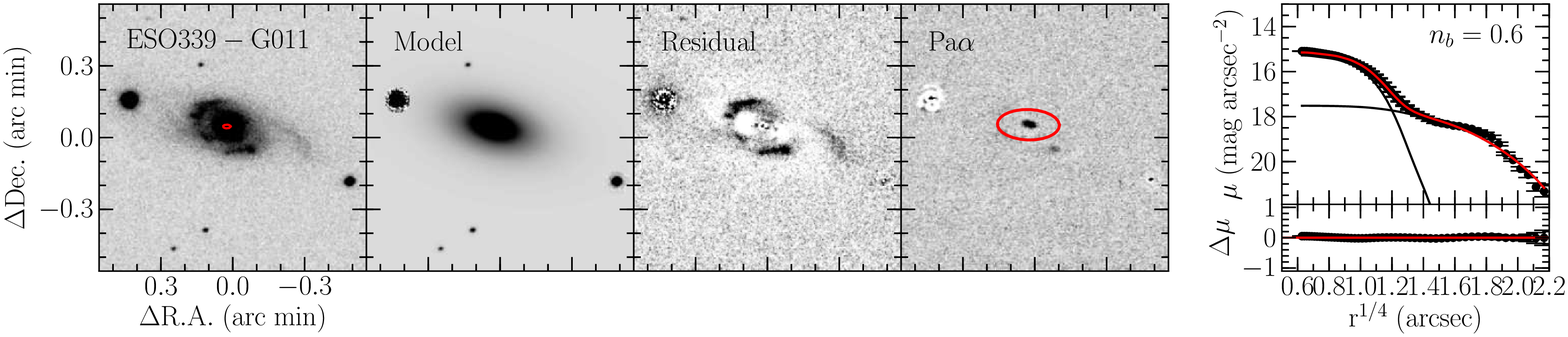}\\
\includegraphics[width=16cm]{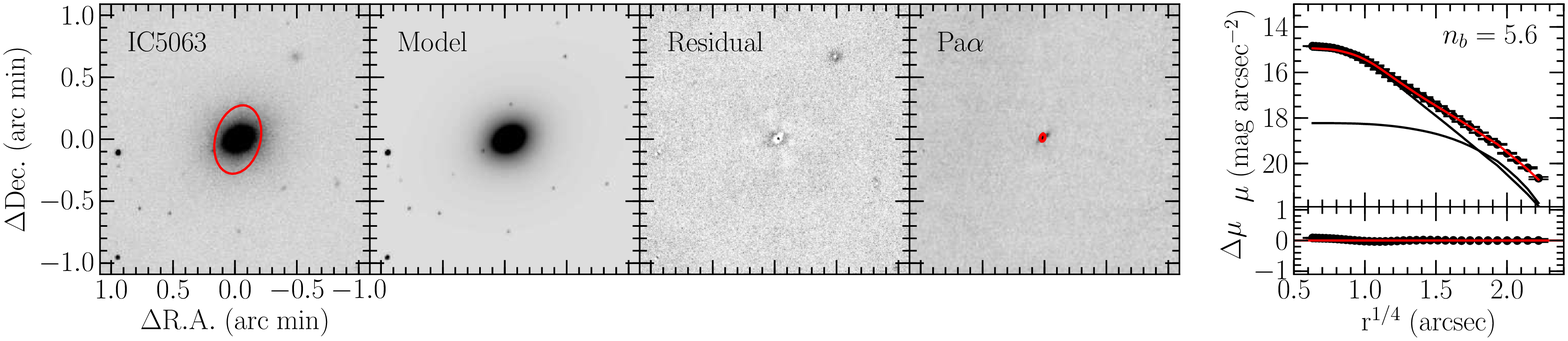}\\
\includegraphics[width=16cm]{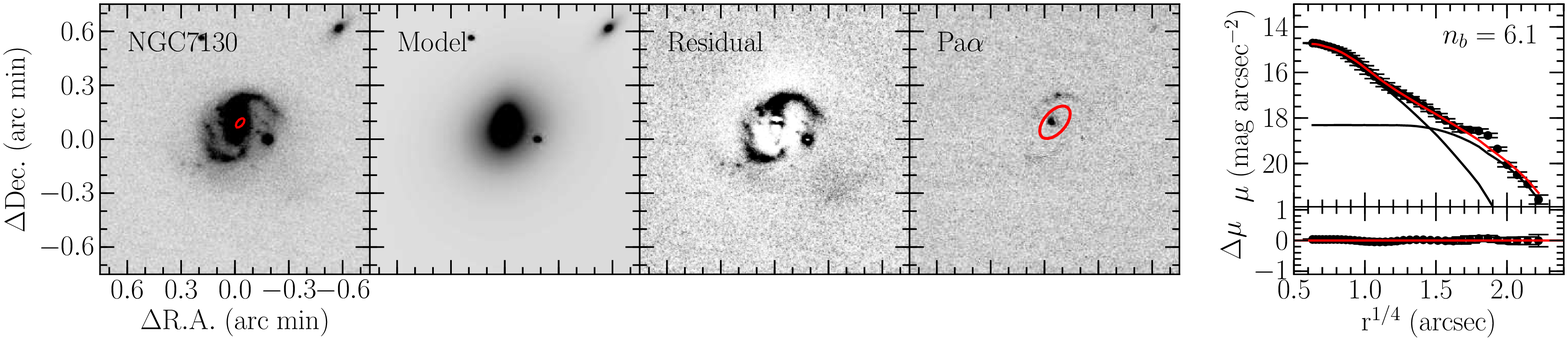}\\
\caption{--Continued. \label{fig:profiles}}
\end{center}
\end{figure*}

\addtocounter{figure}{-1}
\begin{figure*}
\begin{center}
\includegraphics[width=16cm]{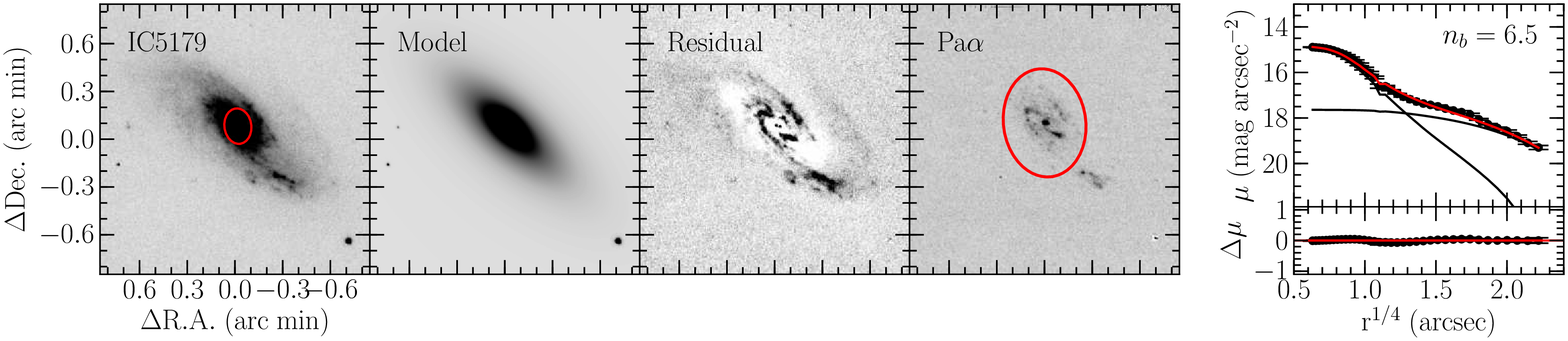}\\
\includegraphics[width=16cm]{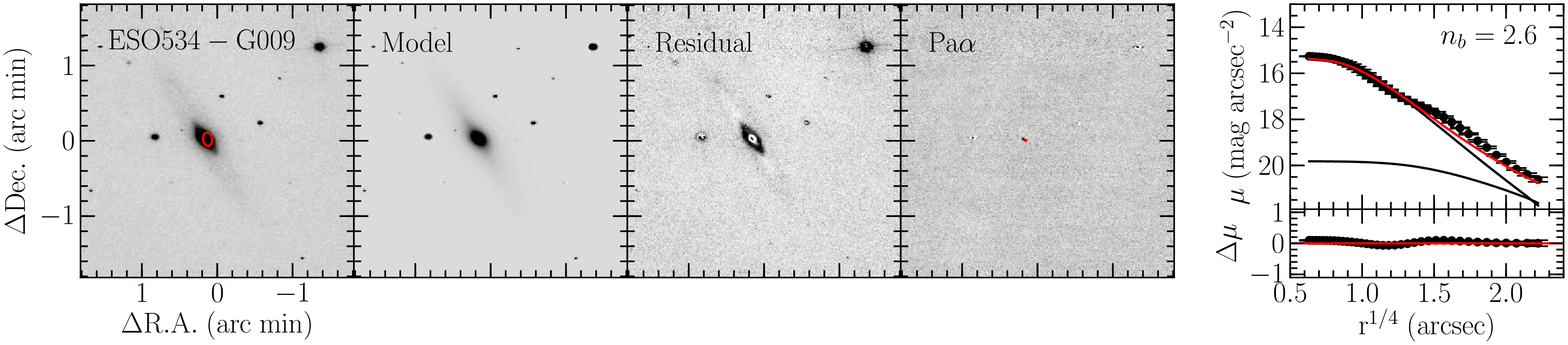}\\
\includegraphics[width=16cm]{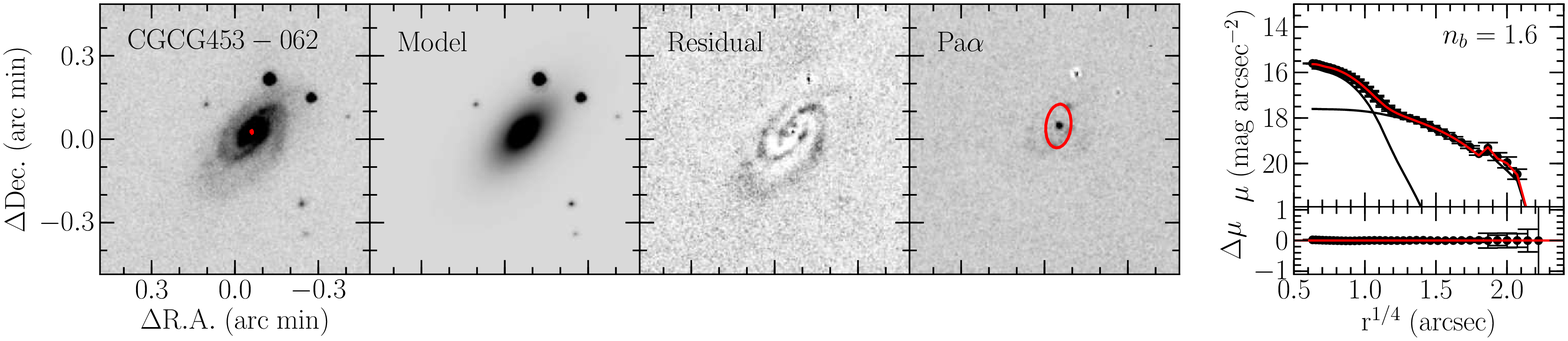}\\
\includegraphics[width=16cm]{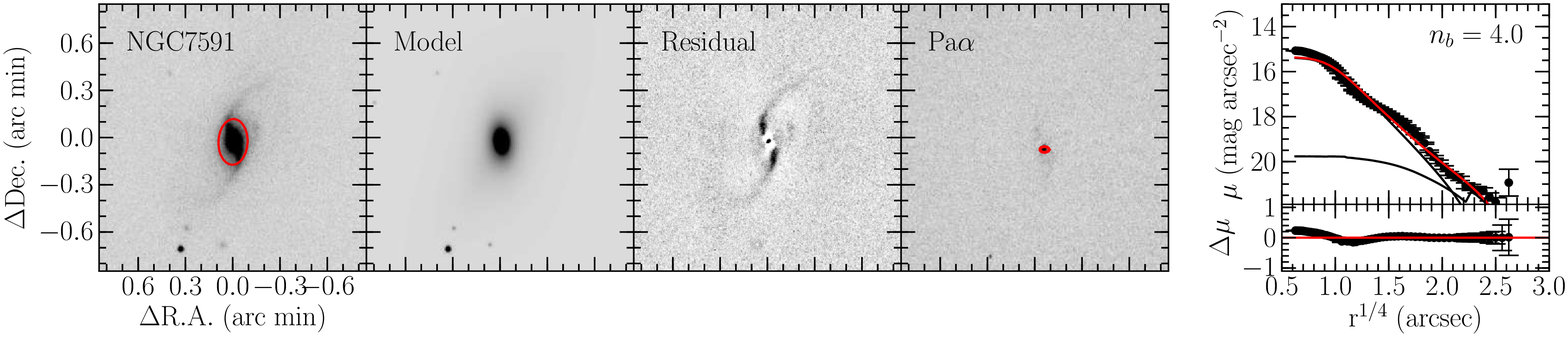}\\
\includegraphics[width=16cm]{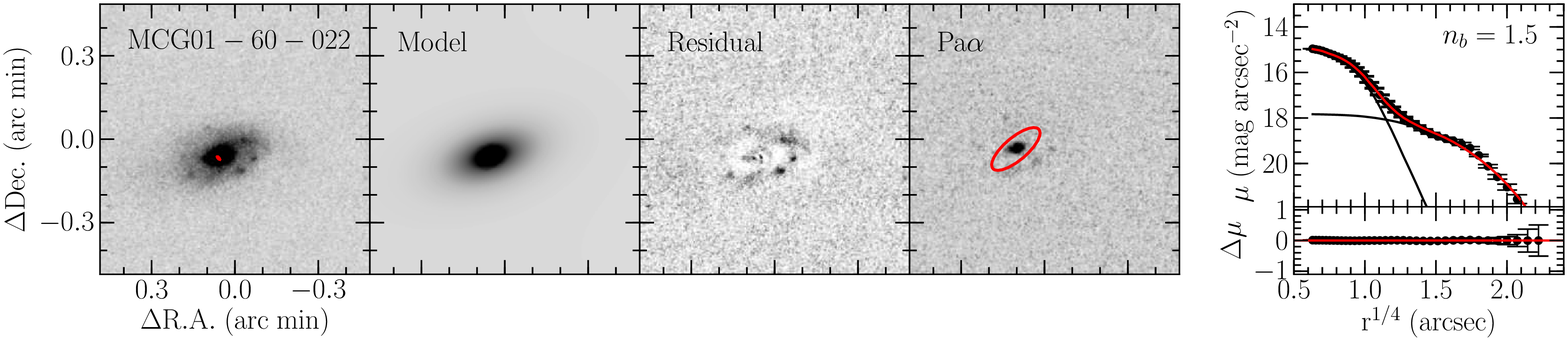}\\
\includegraphics[width=16cm]{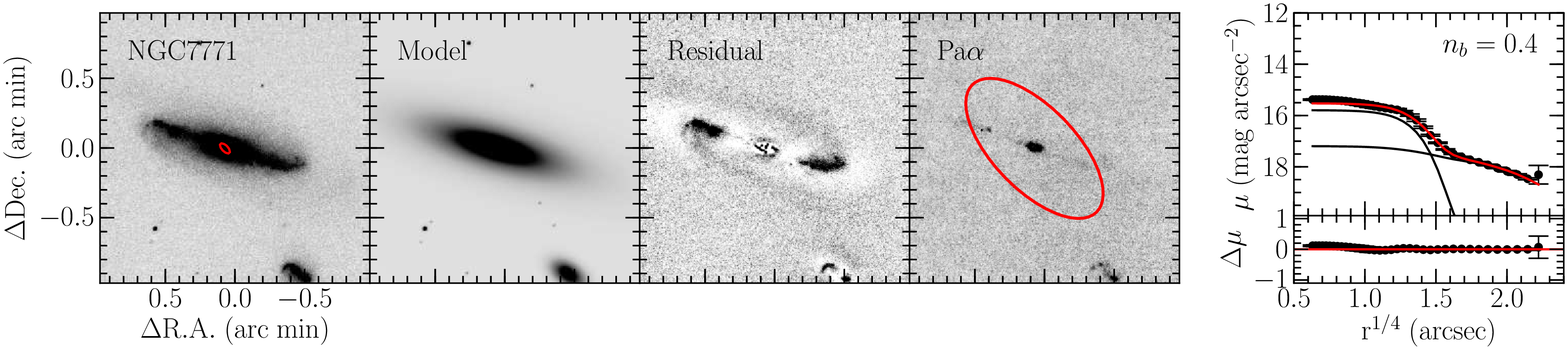}\\
\includegraphics[width=16cm]{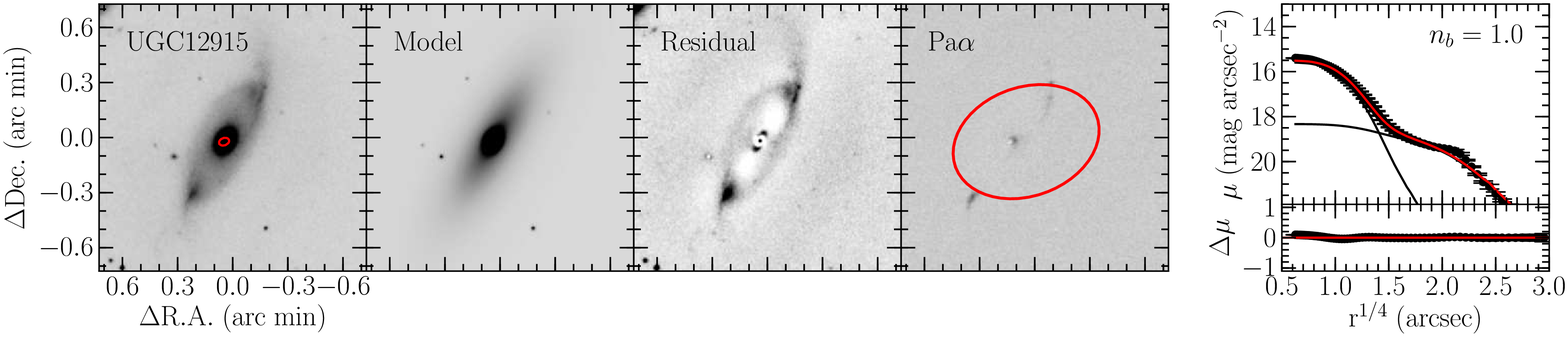}
\caption{--Continued. \label{fig:profiles}}
\end{center}
\end{figure*}

\end{document}